# Analogy and Difference between Gelation and Percolation Process

*Kazumi Suematsu*
Institute of Mathematical Science
Ohkadai 2-31-9, Yokkaichi, Mie 512-1216, JAPAN
Fax: +81 (0) 593 26 8052, E-mail address: suematsu@m3.cty-net.ne.jp

## Summary

It has been verified that the theory of gelation with cyclization effects is in good accord with experimental observations of gel points and gel fractions. Encouraged by this success we scrutinize the prediction limit of the theory through the rigor of the bond percolation theory. Significant disparity is found between the prediction of the gelation theory and that of the percolation theory. To find the reason of the disparity, we re-examine the distribution function of bond animals; the analysis showing that the percolation process differs from real gelations in two points: (i) whereas the real gelation obeys the principle of equireactivity of functional units, the percolation process does not; (ii) the substantial reduction of functionality occurs through the percolation process. These make the lattice model intrinsically different from real chemical processes. As a result, one can not make use of the percolation theory for the purpose of examining the validity of the gelation theory.

## 1. Introduction

For decades, much attention has been focussed on the theory of gelation that takes into account cyclization effects. In a series of papers, the author has reported some closed equations that can be derived under four basic premises. Comparison with experiments carried out subsequently has revealed that the theoretical prediction is in good agreement with the observed values for gel points, critical dilution points and sol-gel lines, in support of the soundness of the theory. Despite those success, it has been suspected that theoretical errors may be hidden behind experimental errors, since the experimental determination of gel points is often difficult and requires many techniques. Thus it seemed essential for the author to scrutinize the predictability limit of the theory from more rigorous points of view. The percolation model seemed to be the best tool for this purpose, since it provides exact results for some particular cases [1]. Hence, in this paper we re-examine the applicability limit of the author's theory of gelation [2] in light of the rigor of condensed matter physics.

Prior to entering the scrutinization of the gelation theory, it will be useful to summarize the background of the theory, which is based on the following four premises:

(i) *The gel point is divided into the two terms*:
$$D_c = D(inter) + D(ring). \tag{1}$$

(ii) *The total ring concentration, $[\Gamma]$, is independent of the initial monomer concentration, C.*

(iii) *Branched molecules behave ideally at $C = \infty$.*

(iv) Assumption I: *Cyclic bonds distribute randomly over all bonds.*

Of these, the most important is the premise (iv) which needs careful treatment, since it is an approximation in contrast to the other three premises and therefore may have some effects on the quality of theoretical prediction. As shown precedently, Assumption I gives the equality $D(inter) = D_{co}$ so that
$$D_c = D_{co} + p_R, \tag{2}$$
with $D_{co}$ being the ideal gel point, and $p_R$ the fraction of cyclic bonds to all possible bonds and can be equated with $D(ring)$ [2]. From eq. (2), all informations about gelation can be obtained: gel points, critical dilution, sol and gel fractions, permanent sol, and so forth. Although so far no serious inconsistency has been found with eq. (2), we must be careful to the extension of the equality. Then let us start the re-examination of the theory.

## 2) Examination of the Theory through the comparison with the Percolation Model

Assumption I directly leads to the equality (2). The examination of Assumption I is therefore equivalent to the examination of equality (2). For the homogeneous $R-A_f$ model of $J = 2$, it follows that $p_R = 2[\Gamma]/fC$ with C being the initial monomer concentration. For the percolation model, $[\Gamma]/C$ can be equated with the number of rings per site. Let $\Omega$ be the number of all clusters. It is well-known that

$$\Omega(D) - \Omega_0(D) = \Gamma(D), \tag{3}$$

the subscript 0 denoting the ideal tree case with no rings. The relationship (3) is obtained naturally from the fact that $\Omega$ decreases through intermolecular reactions alone. For the percolation model, the cluster number per site is $\Omega/M_0$, which has been solved rigorously for the square lattice (SQ). Temperley and Lieb [3], and Ziff [4] showed that, at $D_c$

$$\Omega(D_c)/M_0 = \frac{3\sqrt{3} - 5}{2}, \tag{4}$$

while simulation experiments have shown that $D_c \cong 1/2$ [5]. For SQ, the quantity $\Omega/M_0$ coincides with the number of rings per site $\Gamma/M_0$. It follows from eq. (2) that

$$D_{co} = \frac{1}{2} - \frac{3\sqrt{3} - 5}{4} \cong 0.451. \tag{5}$$

In striking contrast, the Flory formula tells us $D_{co} = 1/(f-1) = 1/3$, since $f = 4$ for SQ. It becomes apparent that significant disparity exists between the result of eq. (5) and the Flory prediction [6]. Such large discrepancy can by no means be explained by the cyclization effect alone.

Let us investigate the reason of the disparity between 0.451 and 1/3. Let $L_n$ be an $n$-cluster, where the subscript $n$ denotes the number of bonds rather than sites, so that $n = 0, 1, 2, \cdots$. The formation of an $n$-cluster on lattices occurs through the reaction of the type:

$$L_{j-1} + L_{n-j} \rightarrow L_n, \tag{c1}$$

while the disappearance occurs through the collision with all other clusters:

$$L_n + L_x \rightarrow L_{n+x+1}. \tag{c2}$$

The formal description for these chemical dynamics is

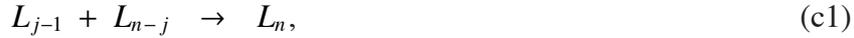
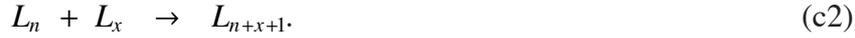

$$\delta n_n = \left( \frac{\frac{1}{2}\sum_{j=1}^{n} \nu_{j-1} n_{j-1} \cdot q_{n-j} - \nu_n n_n}{\frac{1}{2}\sum_{x}^{\infty} \nu_x n_x} \right) \delta u, \tag{6}$$

where $n_x$ denotes the number of $x$-clusters per site, $\delta u$ is unit reaction per site and $\nu_x$ the total number of possible bonds for an $x$-cluster. For smaller clusters, this number $\nu_x$ coincides with the number of unreacted FU's on an $x$-cluster. For larger clusters, however, this isn't the case as becomes clear in the following. The denominator of the right hand side of eq. (6) represents the total number of possible bonds. $q_{n-j}$ represents the probability that a neighboring site of a $(j-1)$-cluster is occupied by an $(n-j)$-cluster, so that the first term of the numerator represents the number of possible bonds for the $n$-cluster formation, and the second term the number of possible bonds for the disappearance of the $n$-cluster.

### Equireactivity Case

For the random reaction of the equireactivity, $q_{n-j}$ is simply given by

$$q_{n-j} = v_{n-j} n_{n-j} \Big/ \sum_{x}^{\infty} v_x n_x. \tag{7}$$

Substituting this into eq. (6) and after rearrangement, one has

$$\delta n_n = \left\{ \frac{\frac{1}{2}\sum_{j=1}^{n} v_{j-1} n_{j-1} \cdot v_{n-j} n_{n-j} - v_n n_n \cdot \sum_{x}^{\infty} v_x n_x}{\frac{1}{2}\left(\sum_{x}^{\infty} v_x n_x\right)^2} \right\} \delta u, \tag{8}$$

which is just of the same form as the kinetic equation for the real system without rings. Eq. (8) is soluble and yields the familiar distribution function for the ideal tree model [2].

## Percolation Process

It is important to stress that for the percolation problem, eq. (6) is generally insoluble, because $q_{n-j}$ is indeterminable. This aspect doesn't appear to have been fully recognized up to present. There are many configurations of an $(n\text{-}j)$-cluster that can not be located at the immediate neighbors of a $(j\text{-}1)$-cluster. In Fig. 1, an example of a 16-cluster with such spatial restriction is illustrated on the $2\,d$ square lattice (SQ). The inner 4 FU's (1, 2, 3, 4) on the 16-cluster have chances to react with $0$-cluster ($\times$) alone, since no other clusters can be located on that site ($\times$) because of the spatial constraint. For those 4 FU's, therefore,

$$q_{n-j} = \begin{cases} 0 & \text{for } n-j \neq 0 \\ 1 & \text{for } n-j = 0 \end{cases}$$

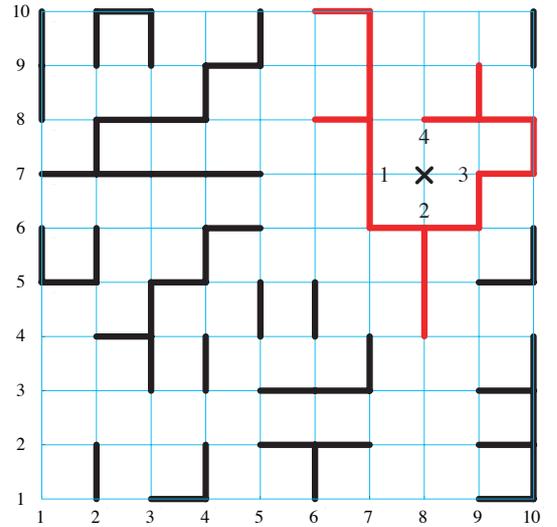

Fig. 1. Bond percolation clusters on SQ.

giving a 17-cluster alone. Any other cluster formation is not possible. There are, of course, many different configurations of the 16-cluster that allow those 4 FU's to react with other clusters. However, in such configurations also, the spatial restriction exists. It is obvious that $q_{n-j}$ cannot generally satisfy eq. (7). Individual FU's on lattices cannot choose their reaction partners at random. This is because of the realization of the so-called steric hindrance in the percolation model.

It is important that the steric hindrance is never manifested in such a manner in actual chemical processes. If such steric hindrance is real, FU's attached to smaller molecules should have larger velocity constants, which does not accord with our experiences. Flory showed earlier [6] citing examples of the esterification of monobasic acids $H(CH_2)_n COOH$ that the reactivity of FU's is independent of the molecular size; in all samples of $3 \leq n \leq \infty$, the velocity constants were found to be invariably $\cong 7.5 \times 10^{-4}$ $[l/gram\ equivalents \cdot sec]$ at 25 °C irrespective of the chain length. Based on these observations, Flory reached the conclusion that every FU has equal opportunity to react with all other FU's, which is the basis of the principle of equireactivity. To date, except for some particular systems, no inconsistency has been found for this principle.

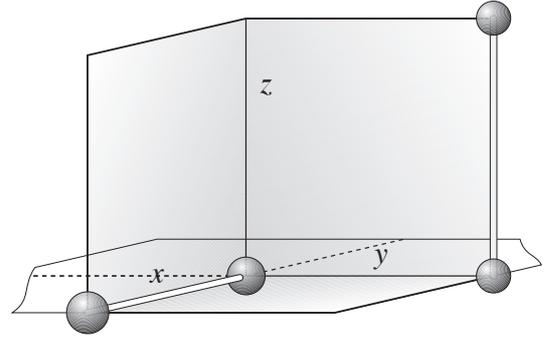

Fig. 2. Formation of a 3-cluster on the cubic lattice.

In the percolation model, because of the incidence of the steric hindrance, there is a deviation from the principle of equireactivity so that the probability $q_{n-j}$ cannot satisfy the equality (7). This means that each configuration of random animals cannot be given the same statistical weight. Thus, since one cannot enumerate all possible reaction pairs over all configurations of all clusters, the probability $q_{n-j}$ is generally indeterminable. In the cluster-distribution point of view, the percolation problem is an insoluble problem.

Putting issues of reality aside, let us calculate, in the following, the animal distribution in a hypothetical system where percolation clusters undergo the equireactivity represented by eq. (7).

**Hypothetical Percolation**

Consider the known percolation process on SQ. But, imagine that every FU has the equal opportunity of reaction with all other clusters. Then $q_{n-j}$ can be equated with eq. (7), so that eq. (6) can be solved successively starting from the *0*-cluster. In SQ, each site has $f = 4$. However, for further extension of our argument, let us use the general notation, $f$. To carry out the calculation, we introduce the notations: *0*: (•), *1*: (•–•), *2I*: (•–•–•), *2L*: (⌐), etc. Using the equality $\delta u = (f/2)\delta D$, one has from eqs. (6) and (7)

$$n_0 = (1-D)^f;$$

$$n_1 = \tfrac{1}{2} fD(1-D)^{2(f-1)};$$

$$n_{2I} = \tfrac{1}{2} fD^2(1-D)^{3f-4};$$

$$n_{2L} = \tfrac{1}{2} f(f-2)D^2(1-D)^{3f-4};$$

so that

$$n_2 = n_{2I} + n_{2L} = \tfrac{1}{2} f(f-1)D^2(1-D)^{3f-4}, \qquad (9)$$

which is exactly equal to the distribution of the ideal tree approximation [2], [6]-[7]:

$$n_n = \omega_n D^n (1-D)^{(f-2)n+f} \quad (n = 0, 1, 2, \ldots),$$

with

$$\omega_n = f \frac{\{(f-1)(n+1)\}!}{(n+1)!\{(f-2)n+f\}!}.$$

The result of eq. (9) is quite reasonable, since those clusters have no rings.

Our purpose is to calculate the distribution of the planar 3U (⊔) cluster, since from this and larger clusters the departure from the ideal distribution is observed. Note that ⊔ can be formed only through the collision of the following two animal pairs:

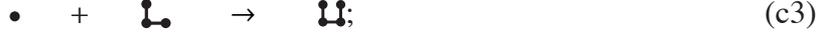  (c3)

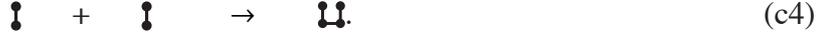  (c4)

Fig. 9

Applying these to eq. (6), we have

$$\tfrac{1}{2}\sum_x^\infty v_x n_x = \tfrac{1}{2} f(1-D);$$

$$\tfrac{1}{2}\sum_{j=1}^n v_{j-1} n_{j-1} q_{n-j} = \begin{cases} 2n_{2L} \cdot f n_0 / f(1-D) & \text{for eq. (c3)} \\ \tfrac{1}{2}\{2(f-2)n_1 \cdot \tfrac{1}{f-2} 2(f-2)n_1 / f(1-D)\} & \text{for eq. (c4)} \end{cases}$$

$$v_n n_n = (4f-7) n_{3U} \qquad \text{for disappearance,}$$

for $D \le D_c$. Substituting these expressions back into eq. (6), one has

$$\delta n_{3U} + \frac{(4f-7)}{1-D} n_{3U}\, \delta D = \tfrac{3}{2} f(f-2) D^2 (1-D)^{4f-6}\, \delta D. \qquad (10)$$

The solution of eq. (10) is

$$n_{3U} = \tfrac{3}{2} f(f-2)(1-D)^{4f-7} \int_0^D D^2 (1-D)\, dD,$$

which giving

$$n_{3U} = \tfrac{3}{8} f(f-2) D^3 (1-D)^{4f-6} + \tfrac{1}{8} f(f-2) D^3 (1-D)^{4f-7}. \qquad (11)$$

Quite identically, we have for all other configurations of 3-clusters $(n_3' = n_3 - n_{3U})$

$$n_3' = \tfrac{1}{6} f(4f^2 - 12f + 11) D^3 (1-D)^{4f-6}. \qquad (12)$$

In terms of the dimension, $n_3 (= n_{3U} + n_3')$ can be written in the form:

$$n_3 = \left\{ d + 16\binom{d}{2} + 32\binom{d}{3} + \left[3\binom{d}{2} + \binom{d}{2}(1-D)^{-1}\right]\right\} D^3 (1-D)^{8d-6}. \qquad (13)$$

Using further the result of eq. (11), the population of the smallest ring, the cyclic 4O (□) cluster, can be calculated by the equation:

$$\delta n_{4O} = \frac{n_{3U} - 4(f-2) n_{4O}}{\tfrac{1}{2} f(1-D)} \delta u, \qquad (14)$$

the solution being

$$n_{4O} = \frac{3}{5}\binom{d}{2}D^4(1-D)^{8d-7} + \frac{2}{5}\binom{d}{2}D^4(1-D)^{8d-8}. \tag{15}$$

We can now compare our results (13) and (15) with the classic work developed in condensed matter physics [8]:

$$n_3(classic) = \left\{d + 16\binom{d}{2} + 32\binom{d}{3} + 4\binom{d}{2}(1-D)^{-1}\right\}D^3(1-D)^{8d-6}; \tag{16}$$

$$n_{4O}(classic) = \binom{d}{2}D^4(1-D)^{8d-8}. \tag{17}$$

The present results (13) and (15) disagree with these classic equations (16) and (17). The reason is comprehensible in a natural fashion from the foregoing derivation. The discrepancy comes from the difference in the estimation of the 3U (⋃) cluster (the fourth term of $n_3$). The classic equations (16) and (17) are obtained just by enumerating total configurations that can be arranged on lattices. However, the percolation process is much more complicated. Whereas the 3U (⋃) cluster is formed through the bond-formation between smaller cluster pairs, it disappears through the reaction with all other clusters. In its formation (c1), there is no configurational constraint, on one hand. In the disappearance (c2), it behaves as if a cluster with the functionality $4f-7$ lower, by 1, than that of the corresponding real molecule $(4f-6)$, on the other hand. The classic calculation neglects this dynamic aspect of the percolation process, which leading to the overestimation of $n_{3U}$.

Much more important is the fact that both our calculation shown above and the classic equations do not describe the true percolation process [9]. Those are solutions for the hypothetical random reaction-system. In real percolation, $q_{n-j}$ is generally indeterminable as mentioned above. Because of the existence of the steric hindrance, the denominator of $q_{n-j}$ should be less than $\sum_x^\infty v_x n_x$: namely, it must be that $q_{n-j} \geq v_{n-j}n_{n-j}/\sum_x^\infty v_x n_x$ for $n_{3U}$. For this reason, the true distribution of $n_{3U}$ should be greater than that shown by eq. (11); i.e., $n_{3U} \geq \left[3\binom{d}{2} + \binom{d}{2}(1-D)^{-1}\right]D^3(1-D)^{8d-6}$. Unfortunately, no experimental verification for this reasoning has been put forth to date. The spatial constraint of configurations is relaxed at high dimensions, so that deviations of eqs. (13) and (16) from the true distribution become negligible with increasing $d$ because of rapid convergence to the ideal distribution of the tree model, but at low dimensions, the deviation is maximal. It turns out that the randomness of the percolation theory means the term, $1/\frac{1}{2}\sum_x^\infty v_x n_x$, in eq. (6), which can be equated with $1/\frac{1}{2}f(1-D)$: In putting bonds on lattices, the percolation process is certainly random, but microscopically, many collisions between animals with particular configurations are forbidden because of the steric hindrance. In the mathematical point of view, putting bonds randomly on lattices is a necessary condition for equireactivity, but not sufficient. In order for the principle of equireactivity to be fulfilled, each FU must have equal opportunities to react with all other FU's. Because of the presence of the configurational constraint, the percolation process cannot satisfy the latter condition, and therefore is a nonrandom process.

We realize that the percolation process differs in two points from real branching reactions: (i) The percolation process does not satisfy the principle of equireactivity; (ii) as the distribution of ⋈ implies, the substantial reduction of functionality occurs throughout the percolation process; such reduction of functionality should increase with increasing $D$ because of the rapid increase in the number of contacts between two FU's on the same cluster. These two effects tend to shift the threshold upward. This is the reason why $D(inter) \cong 0.451$ calculated by eq. (5) so deviates from the ideal value $1/(f-1) = 0.33\cdots$. One can not apply the equality, $D(inter) = 1/(f-1)$, to the percolation process.

## 4. Conclusion

To examine the precision of the prediction, the theory of gelation was applied to the bond percolation. The analysis showed that because of the intrinsic difference in reaction mechanism between the real gelation and the percolation process, one cannot make use of the percolation theory to examine the validity of the polymer theory, especially at low dimensions.